\documentclass[11pt,showpacs, preprintnumbers,superscriptaddress,amsmath,amssymb,nofootinbib]{revtex4}
\usepackage{graphicx}
\usepackage{dcolumn}
\usepackage{bm}
\usepackage{amssymb}
\usepackage{amsmath}
\usepackage{epsfig}    
\usepackage{color}
\usepackage{slashed}
\usepackage{hhline}
\def\be{\begin{equation}}
\def\ee{\end{equation}}
\newcommand{\bea}{\begin{eqnarray}}
\newcommand{\eea}{\end{eqnarray}}
\newcommand{\nn}{\nonumber}

\numberwithin{equation}{section}
\usepackage{hyperref}

\begin{document}


\title{Zee model in a non-holomorphic modular $A_4$ symmetry}

\author{Takaaki Nomura}
\email{nomura@scu.edu.cn}
\affiliation{College of Physics, Sichuan University, Chengdu 610065, China}

\author{Hiroshi Okada}
\email{hiroshi3okada@htu.edu.cn}
\affiliation{Department of Physics, Henan Normal University, Xinxiang 453007, China}

\date{\today}

\begin{abstract}
We study a Zee model in a non-holomorphic modular $A_4$ flavor symmetry in which we find good predictions in both the cases of normal and inverted hierarchy. Parameter reduction on neutrino sector occurs due to large mass hierarchies between charged-leptons mass eigenvalues and new singly-charged bosons in addition to this flavor symmetry. As a result, we have two complex free parameters including modulus $\tau$. We show several predictions in terms of verifiable observables such as Dirac CP, Majorana phases, sum of the neutrino masses, and the effective mass for neutrino double beta decay in addition to demonstrating allowed regions for our input parameters.
 \end{abstract}

\maketitle

\newpage

\section{Introduction}
Searching for plausible scenario for neutrino sector would be important to understand particle physics beyond the standard model (BSM).
Since supersymmetry(SUSY) is unlikely to exist at low energy scale that can reach at our current experiments $\sim$TeV, 
it would be promising to work on non-supersymmetric theory as a first step.
Recently, a group of "Qu" and "Ding" successfully constructed a non-holomorphic modular flavor symmetries that can still work on non-supersymmetric theory~\cite{Qu:2024rns}. 
Thanks to their big efforts, 
more varieties of scenarios have been taken in consideration~\cite{Ding:2024inn, Li:2024svh, Nomura:2024atp, Nomura:2024vzw}.
Zee model~\cite{Zee:1980ai} is one of the attractive scenarios to generate the non-vanishing neutrino masses radiatively since it is expected to be detected at TeV scale.~\footnote{We should also mention Ma model that is the first neutrino model at one-loop including dark matter candidate~\cite{Ma:2006km} although Zee model does not have dark matter.}

In this paper, we apply the non-holomorphic modular $A_4$ flavor symmetry for Zee model and construct the model as minimum as possible.
Thanks to non-holomorphic features, we have drastically  reduced our free parameter compared to our previous scenario under the holomorphic modular $A_4$ symmetry~\cite{Nomura:2021pld}.
As a result, we obtain good predictions in terms of verifiable observables such as Dirac CP, Majorana phases, sum of the neutrino masses, and the effective mass for neutrino double beta decay for both the cases of normal and inverted hierarchy performing $\chi$ square analysis.

This paper is organized as follows.
In Sec.~II, we review our minimum Zee model constructing the renormalizable valid Lagrangian, Higgs potential, charged-lepton mass matrix, and active neutrino mass matrix.
Then, we numerically fix the Higgs masses and mixings so that our analysis makes it simpler.
After discussing the charged-lepton sector, we formulate the neutrino sector. 
In Sec.~III, we perform $\chi$ square analysis and show some predictions for normal and inverted hierarchies.
We have conclusions and discussion in Sec.~IV.

\begin{center} 
\begin{table}[tbh!]
\begin{tabular}{|c||c|c||c|c|c|}\hline\hline  
  & \multicolumn{2}{c|}{Leptons} & \multicolumn{3}{c|}{Bosons}   \\ \hline \hline
& ~$ \overline{L_L}$~& ~${\ell}_R$~& ~$\Phi$~& ~$\Phi'$~& ~$S^-$~       \\\hline\hline 
$SU(2)_L$ & $\bm{2}$  & $\bm{1}$   & $\bm{2}$ & $\bm{2}$ & $\bm{1}$   \\\hline 
$U(1)_Y$   & $-\frac12$ & $1$    & $\frac12$& $\frac12$& $-1$    \\\hline
 $A_4$ & $3$ & $3$ & $1$ & $1$ & $1$         \\ \hline
$-k_I$ & $-1$ & $+1$& $0$& $-2$& $+2$    \\
\hline
\end{tabular}
\caption{Field contents and  their charge assignments in Zee model under $SU(2)_L\times U(1)_Y\times A_4$ where $-k_I$ is the number of modular weight.}
\label{tab:1}
\end{table}
\end{center}

 \section{Model setup}
{The field contents in our model setup is exactly the same as Zee model, i.e. two Higgs doublets $\Phi,\ \Phi'$ plus singly-charged boson $S^-$ are introduced.}
The charge assignments under modular $A_4$ are shown in Tab~\ref{tab:1}. 
The assignments are chosen so that the model is as minimum as possible where we define $\overline{L_L}\equiv[\overline{L_{L_e}} ,\overline{L_{L_\mu}} ,\overline{L_{L_\tau}} ]^T$, $\ell_R\equiv[e_R,\mu_R,\tau_R]^T$ as flavor eigenstates.
Then, the renormalizable Lagrangian for lepton sector is found as follows: 
\begin{align}
 - {\cal L}_\ell & = 
a_e \left[\overline{L_{L_e}} e_R+\overline{L_{L_\mu}} \tau_R+\overline{L_{L_\tau}} \mu_R\right] \Phi
 \nn \\
& +b_e \left[ y_1(2\overline{L_{L_e}} e_R-\overline{L_{L_\mu}} \tau_R-\overline{L_{L_\tau}} \mu_R)
+y_2(2\overline{L_{L_\mu}} \mu_R-\overline{L_{L_e}} \tau_R-\overline{L_{L_\tau}} e_R) \right.  \nn\\
&\left. +y_3(2\overline{L_{L_\tau}} \tau_R-\overline{L_{L_e}} \mu_R-\overline{L_{L_\mu}} e_R \right] \Phi
\nn\\
& +c_e \left[y_1( \overline{L_{L_\mu}} \tau_R-\overline{L_{L_\tau}} \mu_R)
+y_2( -\overline{L_{L_e}} \tau_R+\overline{L_{L_\tau}} e_R)
+y_3( \overline{L_{L_e}} \mu_R-\overline{L_{L_\mu}} e_R)\right] \Phi
\nn\\
& +a' \left[ f_1(2\overline{L_{L_e}} e_R-\overline{L_{L_\mu}} \tau_R-\overline{L_{L_\tau}} \mu_R)
+f_2(2\overline{L_{L_\mu}} \mu_R-\overline{L_{L_e}} \tau_R-\overline{L_{L_\tau}} e_R) \right.  \nn\\
&\left. +f_3(2\overline{L_{L_\tau}} \tau_R-\overline{L_{L_e}} \mu_R-\overline{L_{L_\mu}} e_R \right] \Phi' \nn\\
& +b'_k \left[f_1( \overline{L_{L_\mu}} \tau_R-\overline{L_{L_\tau}} \mu_R)
+f_2( -\overline{L_{L_e}} \tau_R+\overline{L_{L_\tau}} e_R)
+f_3( \overline{L_{L_e}} \mu_R-\overline{L_{L_\mu}} e_R)\right] \Phi' \nonumber \\
&+ a_s \left[y_1( \overline{L_{L_\mu}} L^C_{L_\tau}-\overline{L_{L_\tau}} L^C_{L_\mu})
+y_2( -\overline{L_{L_e}} L^C_{L_\tau} +\overline{L_{L_\tau}} L^C_{L_e})
+y_3( \overline{L_{L_e}} L^C_{L_\mu}-\overline{L_{L_\mu}} L^C_{L_e})\right] S^- +{\rm h.c.}, 
\label{yukawa}
\end{align}
where we define $Y_3^{(0)} = [y_1,y_2,y_3]$ and  $Y_3^{(2)} = [f_1,f_2,f_3]$~\cite{Qu:2024rns}. 
{
\subsection{Higgs sector}
The Higgs sector of the model is the same as the one in the Zee model.
The relevant scalar sector is given by
\begin{align}
  {\cal V} &= \mu_1^2 |\Phi'|^2 + \mu^2_2 |\Phi|^2  - ( \mu_{3}^2 \Phi^\dagger \Phi' + h.c. ) + \mu_S^2 |S^-|^2 + \mu (\Phi^T \cdot \Phi')S^- \nn\\
&+ \frac{\lambda_1}{2} |\Phi'|^4+ \frac{\lambda_2}{2} |\Phi|^4 + \lambda_S |S|^4 + \lambda_{3}|\Phi|^2|\Phi'|^2 + \lambda_{4}|\Phi^\dag \Phi'|^2  + \frac{\lambda_{5}}{2}((\Phi^\dag \Phi')^2 +h.c.)  \nn\\
&+ \lambda_{\Phi S}|\Phi|^2|S|^2+  \lambda_{\Phi' S}|\Phi'|^2|S|^2
  ,\label{Eq:pot}
\end{align}
where all the above parameters except $\mu_2^2$, $\lambda_2$, and $\mu$ include $1/(\tau-\tau^*)^{k_I}$ factor, and $\mu^2_{3}$ and $\lambda_5$ respectively contain modular form $Y^{(2)}_1$ and $Y^{(4)}_1$ in order to be invariant under the modular $A_4$ symmetry;
these extra factors are absorbed into the parameters after $\tau$ value is fixed.
Here we introduce the Higgs basis $(H,H')$ as
\begin{equation}
\begin{pmatrix} \Phi' \\ \Phi \end{pmatrix} = 
\begin{pmatrix} c_\beta & -s_\beta \\  s_\beta & c_\beta \end{pmatrix}
\begin{pmatrix} H \\ H' \end{pmatrix}
\end{equation}
where $c_\beta(s_\beta) \equiv \cos \beta (\sin \beta)$ with mixing angle $\beta$ defined by the ratio of vacuum expectation values (VEVs) as $\tan \beta = \langle \Phi \rangle/\langle \Phi' \rangle$.
Then $H$ and $H'$ are written as follows:
\begin{equation}
H=
\begin{pmatrix} w^+ \\ \frac{v + \tilde{h}+i z }{\sqrt2} \end{pmatrix} ,\quad
H'=
\begin{pmatrix} h'^+ \\ \frac{ \tilde{h}'+i A }{\sqrt2} \end{pmatrix} ,
\end{equation} 
where $v\approx 246$ GeV is the VEV in the Higgs basis after the spontaneous symmetry breaking, $z$ is absorbed by the neutral gauge boson of the SM $Z$, and  $w^+$ is absorbed by the charged gauge boson of the SM $W^+$.
As in the two Higgs doublet model (THDM) the mass eigenstates for the CP even physical bosons are written by
\begin{align}
\begin{pmatrix} \tilde{h} \\ \tilde{h}' \end{pmatrix} = 
\begin{pmatrix} c_{\alpha - \beta} & -s_{\alpha - \beta} \\  s_{\alpha - \beta} & c_{\alpha - \beta} \end{pmatrix}
\begin{pmatrix} H \\ h \end{pmatrix},
\end{align}
where the mixing angle $\alpha$ can be expressed in terms of parameters in the potential, and mass eigenvalues are $m_h$ and $m_H$.
In our analysis, we consider the alignment limit of $\sin (\beta - \alpha) =1$ to avoid experimental constraints associated with Higgs boson measurements.
For simplicity we also choose parameters in the potential so that $\tan \beta \gg 1$ and $\Phi \simeq H$ is the SM like Higgs field.
Thus we approximate as $\langle \Phi \rangle \simeq v/\sqrt{2}$ and neglect the SM fermion mass terms from VEV of $\Phi' \simeq H'$.

The mass eigenvalue of CP odd one is given by
\begin{align}
m_{A}^2&= \frac{\mu_3^2}{s_\beta c_\beta} - v^2 \lambda_5.
\end{align}
The mass matrix of singly-charged bosons is given by
\begin{align}
M^2_C
= 
\begin{pmatrix}
\frac{\mu_3^2}{s_\beta c_\beta} - \frac{v^2}{2} (\lambda_4 + \lambda_5)  & \frac{\mu v}{\sqrt2} \\
\frac{\mu v}{\sqrt2} & \mu^2_S +\frac{v^2 s^2_\beta}{2}\lambda_{\Phi S} + \frac{v^2 c^2_\beta}{2}\lambda_{\Phi' S}  \end{pmatrix}.
\end{align}
Here, we presume that the singly-charged bosons are almost diagonal assuming
 $\mu \ll v$. Thus, we consider that the neutrino mass matrix is found through the mass insertion approximation as shown below.
 Therefore, 
 \begin{align}
m^2_{h'}&\approx \frac{\mu_3^2}{s_\beta c_\beta} - \frac{v^2}{2} (\lambda_4 + \lambda_5),\\
m^2_{s}&\approx  \mu^2_S +\frac{v^2 s^2_\beta}{2}\lambda_{\Phi S} + \frac{v^2 c^2_\beta}{2}\lambda_{\Phi' S}.
\end{align}

}

\subsection{Charged-lepton mass matrix}
After the spontaneous electroweak symmetry breaking,
the charged-lepton mass matrix $M_e$ is given by
\begin{align}
&M_e = \frac{v}{\sqrt2} \left[
a_e \begin{pmatrix}
1 & 0 & 0 \\ 
 0 &  0 &1 \\ 
0 & 1 & 0 \\ 
\end{pmatrix}
+ b_e
\begin{pmatrix}
 2y_1 & - y_3 &- y_2 \\ 
- y_3 & 2 y_2& -y_1 \\ 
-y_2 & - y_1 & 2 y_3 \\ 
\end{pmatrix}
+ c_e
\begin{pmatrix}
 0 &  y_3 &- y_2 \\ 
- y_3 & 0 & y_1 \\ 
y_2 & - y_1 & 0 \\ 
\end{pmatrix} \right],
 \label{massmat}
\end{align}
where $a_e,b_e,c_e$ are real for simplicity. 
Then, the charged-lepton mass matrix is diagonalized by a bi-unitary mixing matrix as $D_e\equiv{\rm diag}(m_e,m_\mu,m_\tau)=V^\dag_{eL} M_e V_{eR}$.
These three parameters are used in order to fit the mass eigenvalues of charged-leptons by solving the following three relations:
\begin{align}
&{\rm Tr}[M_e M_e^\dag] = |m_e|^2 + |m_\mu|^2 + |m_\tau|^2,\\
&{\rm Det}[M_eM_e^\dag] = |m_e|^2  |m_\mu|^2  |m_\tau|^2,\\
&({\rm Tr}[M_eM_e ^\dag])^2 -{\rm Tr}[(M_e M_e^{\dag})^2] =2( |m_e|^2  |m_\nu|^2 + |m_\mu|^2  |m_\tau|^2+ |m_e|^2  |m_\tau|^2 ).
\end{align}

\subsection{Active neutrino mass matrix} 
\label{neut}
The active neutrino mass matrix is given at one-loop level via the following Lagrangian in terms of mass eigenstates of charged-leptons and singly-charged-bosons:
\begin{align}
a' \overline{\nu_L} f V_{e_R} \ell_R h'^+
+a_s\overline{\ell_L} V_{e_L}^\dag  y^T \nu^c_L  S^- +\frac{\mu v}{\sqrt2} h'^- S^+
 +{\rm h.c.},
\label{yukawa}
\end{align}
where $f$ and $y$ are respectively given by
\begin{align}
f& = 
\begin{pmatrix}
 2f_1 & - f_3 &- f_2 \\ 
- f_3 & 2 f_2& -f_1 \\ 
-f_2 & - f_1 & 2 f_3 \\ 
\end{pmatrix} + 
\tilde b'_k
\begin{pmatrix}
 0 &  f_3 &- f_2 \\ 
- f_3 & 0 & f_1 \\ 
f_2 & - f_1 & 0 \\ 
\end{pmatrix},
\\
y& = 
\begin{pmatrix}
 0 &  y_3 &- y_2 \\ 
- y_3 & 0 & y_1 \\ 
y_2 & - y_1 & 0 \\ 
\end{pmatrix},
\end{align}
where $\tilde b_k\equiv b_k/a'$.
The neutrino mass matrix is found at one-loop level as follows:
\begin{align}
& (m_{\nu})_{ij}  \approx \frac{a_s a'}{(4 \pi)^2}\frac{v\mu}{\sqrt2 m^2_{h'}}\frac{\ln r}{1-r}
 \sum_{a=1} \left(F_{ja} D_{\ell_a}  Y^T_{aj} +  Y_{ia} D_{\ell_a}  F^T_{aj} \right) ,
\end{align}
where $r\equiv m^2_s/m^2_{h'}$, $F\equiv f V_{eR}$, $Y\equiv y V^*_{eL}$, and
the loop function does not depend on the mass of charged-leptons since we assume  $m_{e,\mu,\tau} \ll m_{h'},\ m_s$.
We define the overall factor as follows: $\kappa\equiv  \frac{a_s a'}{(4 \pi)^2}\frac{v\mu}{\sqrt2 m^2_{h'}}\frac{\ln r}{1-r}$,
therefore $m_{\nu}\equiv \kappa \tilde m_\nu$.
Then, $\tilde m_\nu$ is diagonalized by a unitary matrix $U_\nu$ as $U_\nu^T \tilde m_\nu U_\nu =\tilde D_\nu$ with $\tilde D_\nu = {\rm diag}[\tilde D_{\nu_1},\tilde D_{\nu_2},\tilde D_{\nu_3}]$,
and the Pontecorvo-Maki-Nakagawa-Sakata unitary matrix $U_{PMNS}$ is defined by $V_{eL}^\dag U_\nu$. 
The observed atmospheric mass squared difference $\Delta m^2_{atm}$ is given by 
\begin{align}
&{\rm NH}:\ \Delta m^2_{atm}= |\kappa|^2 (\tilde D^2_{\nu_3} - \tilde D^2_{\nu_1}),\\
&{\rm IH}:\ \Delta m^2_{atm}= |\kappa|^2 (\tilde D^2_{\nu_2} - \tilde D^2_{\nu_3}),
\end{align}
where NH(IH) represents normal(inverted) hierarchy.
The solar mass squared difference $\Delta m^2_{sol}$ is then given by
\begin{align}
\Delta m^2_{sol}= |\kappa|^2 (\tilde D^2_{\nu_2} - \tilde D^2_{\nu_1}).
\end{align}
Finally, the effective mass for neutrino double beta decay is given by
\begin{align}
\langle m_{ee}\rangle = |\kappa| \left|\tilde D_{\nu_1} c^2_{12} c^2_{13}+\tilde D_{\nu_2} s^2_{12} c^2_{13}e^{i\alpha_{21}}
+\tilde D_{\nu_3} s^2_{13}e^{i(\alpha_{31}-2\delta_{CP})} \right|.
\end{align}

A current KamLAND-Zen data~\cite{KamLAND-Zen:2024eml}. provides measured observable in future and 
its upper bound is given by $\langle m_{ee}\rangle<(36-156)$ meV at 90 \% confidence level.
%
The minimal cosmological model
$\Lambda$CDM $+\sum D_{\nu}$ provides upper bound on $\sum D_{\nu}\le$ 120 meV~\cite{Vagnozzi:2017ovm, Planck:2018vyg}.
Moreover, recently combination of DESI and CMB data gives more stringent upper bound on this bound;
$\sum D_{\nu}\le$ 72 meV~\cite{DESI:2024mwx}. 

\section{Numerical analysis}
In this section we perform $\chi$ square analysis using data from NuFit6.0~\cite{Esteban:2024eli}
where we have adopt five reliable observables; three mixings, two mass square differences, for the analysis. 
The green points represents the interval of $1\sigma-2\sigma$, yellow one $2\sigma-3\sigma$, and red one $3\sigma-5\sigma$.
Our input parameter, which is complex, is randomly selected within the following range:
\begin{align}
&|\tilde b'_k | \in [10^{-5},10^5],
\end{align}
where we work on the fundamental region of $\tau$.

\subsection{NH}


\begin{figure}[t]
  \includegraphics[width=77mm]{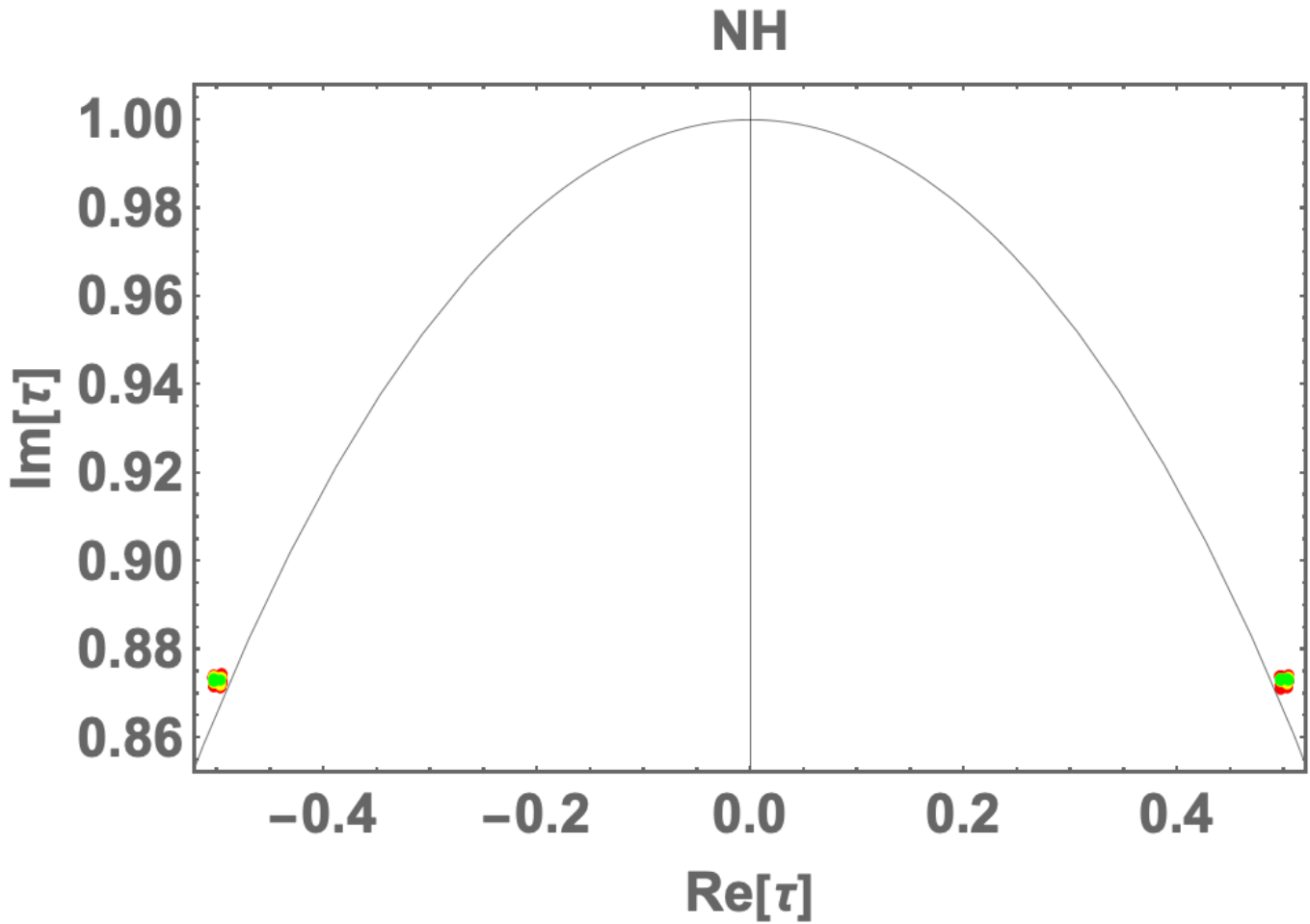}
  \caption{Allowed region for real $\tau$ and imaginary $\tau$ in NH.}
  \label{fig:tau_nh}
\end{figure}
In Fig.~\ref{fig:tau_nh}, we show the allowed region of $\tau$, and find that the allowed region is located at nearby $\tau=e^{2\pi i/3}$.
\begin{figure}[t]
  \includegraphics[width=77mm]{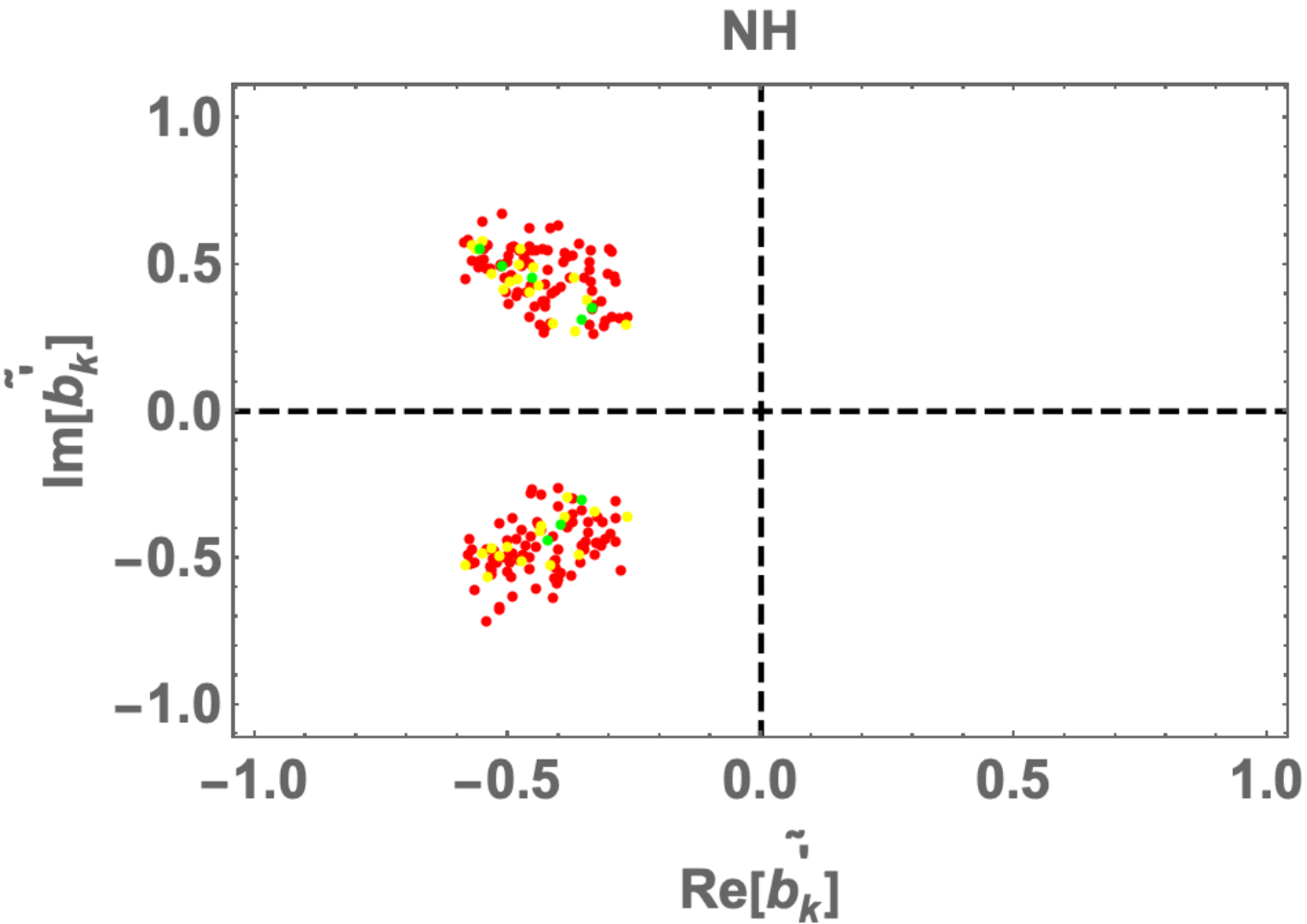}
  \caption{Allowed region for real $\tilde b'_k$ and imaginary $\tilde b'_k$ in NH.}
  \label{fig:bk_nh}
\end{figure}
In Fig.~\ref{fig:bk_nh}, we also show the allowed region of $\tilde b'_k$, and find that the allowed region is about
${\rm Re}[\tilde b'_k]=[-0.6,-0.2]$ and $|{\rm Im}[\tilde b'_k]|=[0.2,0.6]$.

\begin{figure}[t]
  \includegraphics[width=77mm]{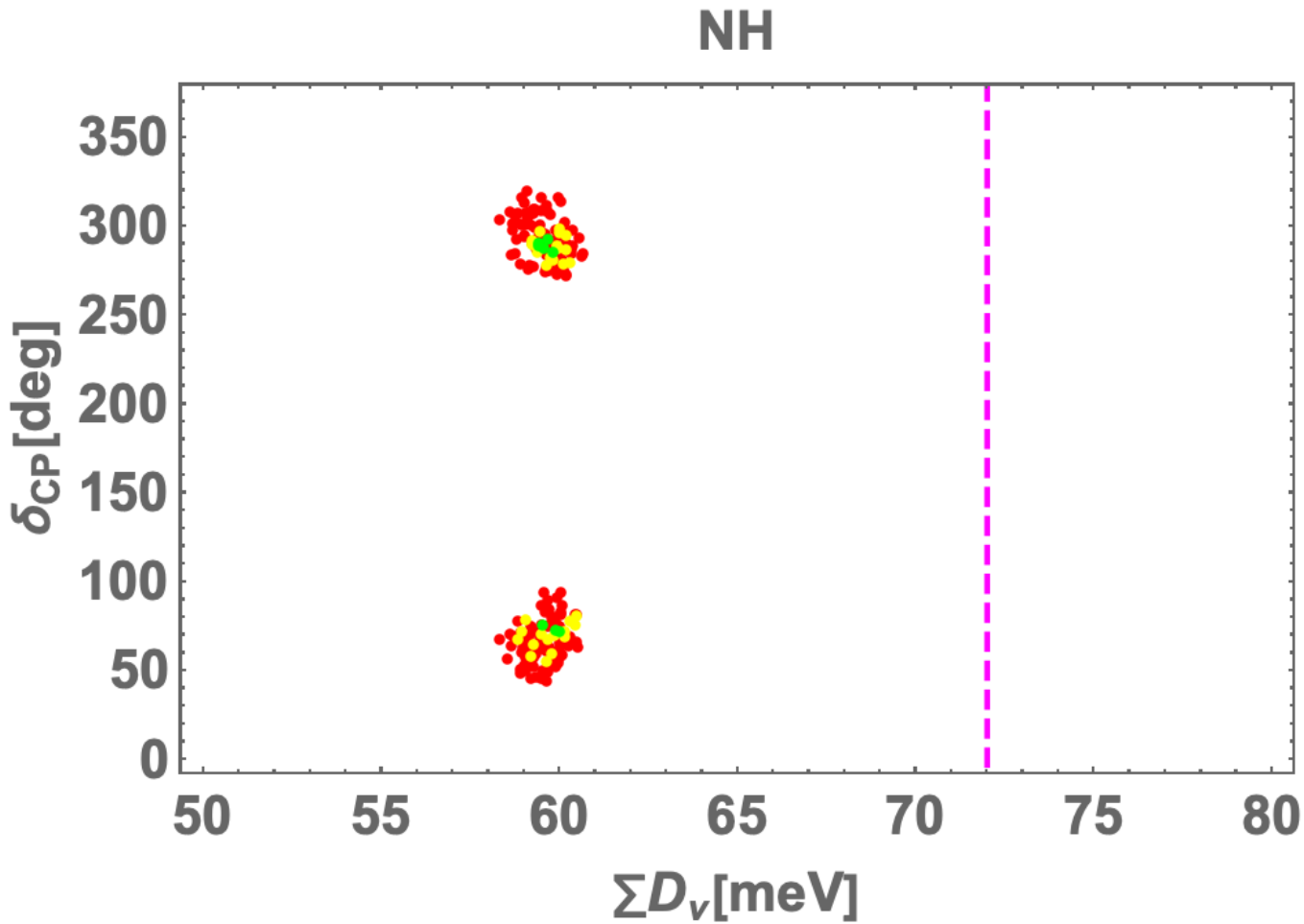}
    \includegraphics[width=77mm]{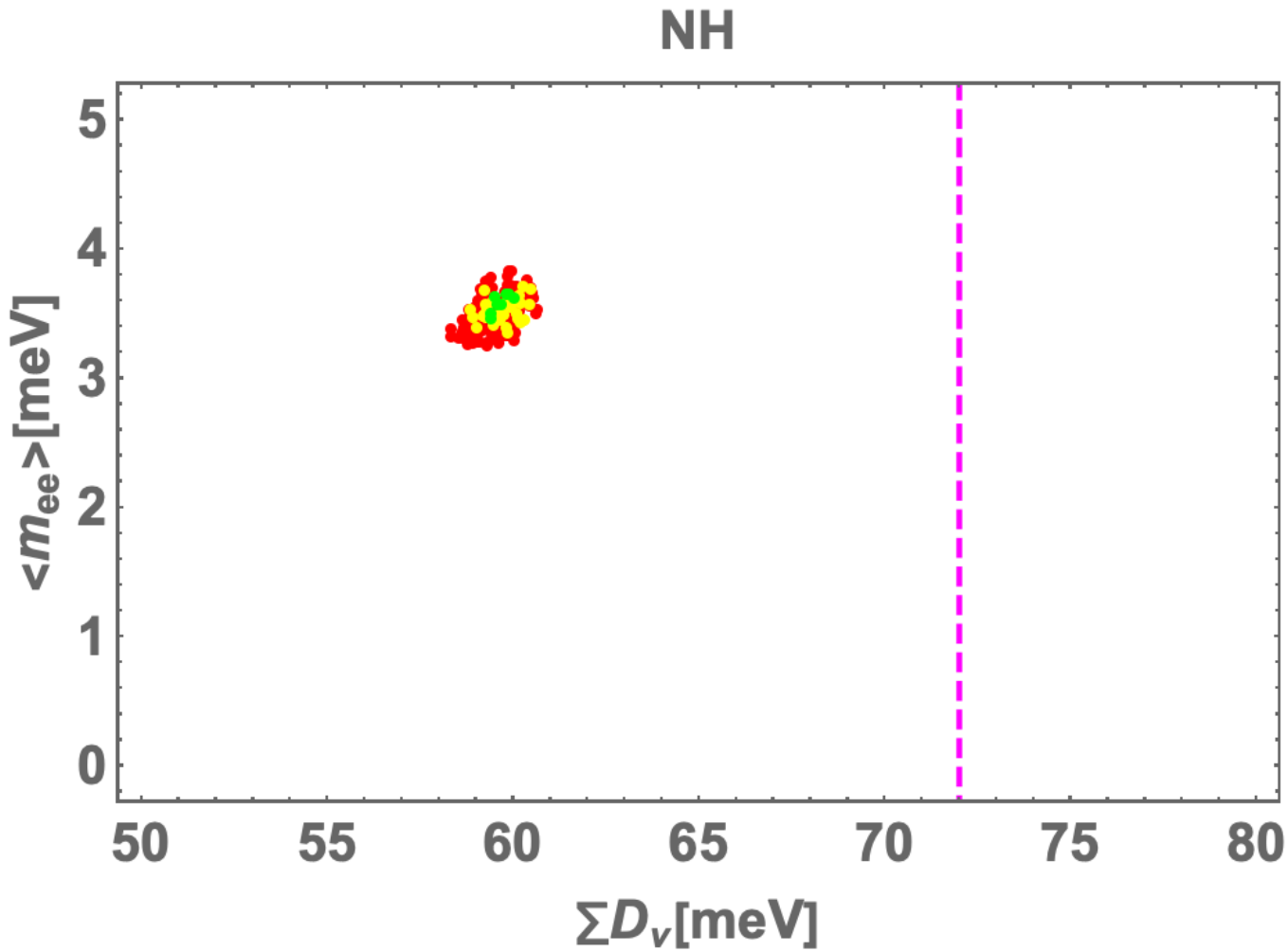}
  \caption{Allowed regions for  $\delta_{CP}$ deg (left) and $\langle m_{ee}\rangle$ meV (right) in terms of $\sum D_\nu$ meV in NH. The vertical magenta dotted line is upper bound on results of Planck+DESI~\cite{DESI:2024mwx} $\sum D_\nu\le $72 meV. }
  \label{fig:sum-dcp_nh}
\end{figure}
In Fig.~\ref{fig:sum-dcp_nh}, we show the allowed  region for $\delta_{CP}$ deg (left) and $\langle m_{ee}\rangle$ meV (right) in terms of $\sum D_\nu$ meV and 
find $\delta_{CP} =[30-100, 270-330]$ deg,  $\sum D_\nu =[58-61]$ meV, and $\langle m_{ee}\rangle =[3.2-4.0]$ meV.
 The vertical magenta dotted line is upper bound on results of Planck+DESI~\cite{DESI:2024mwx} $\sum D_\nu\le $72 meV.

\begin{figure}[t]
  \includegraphics[width=77mm]{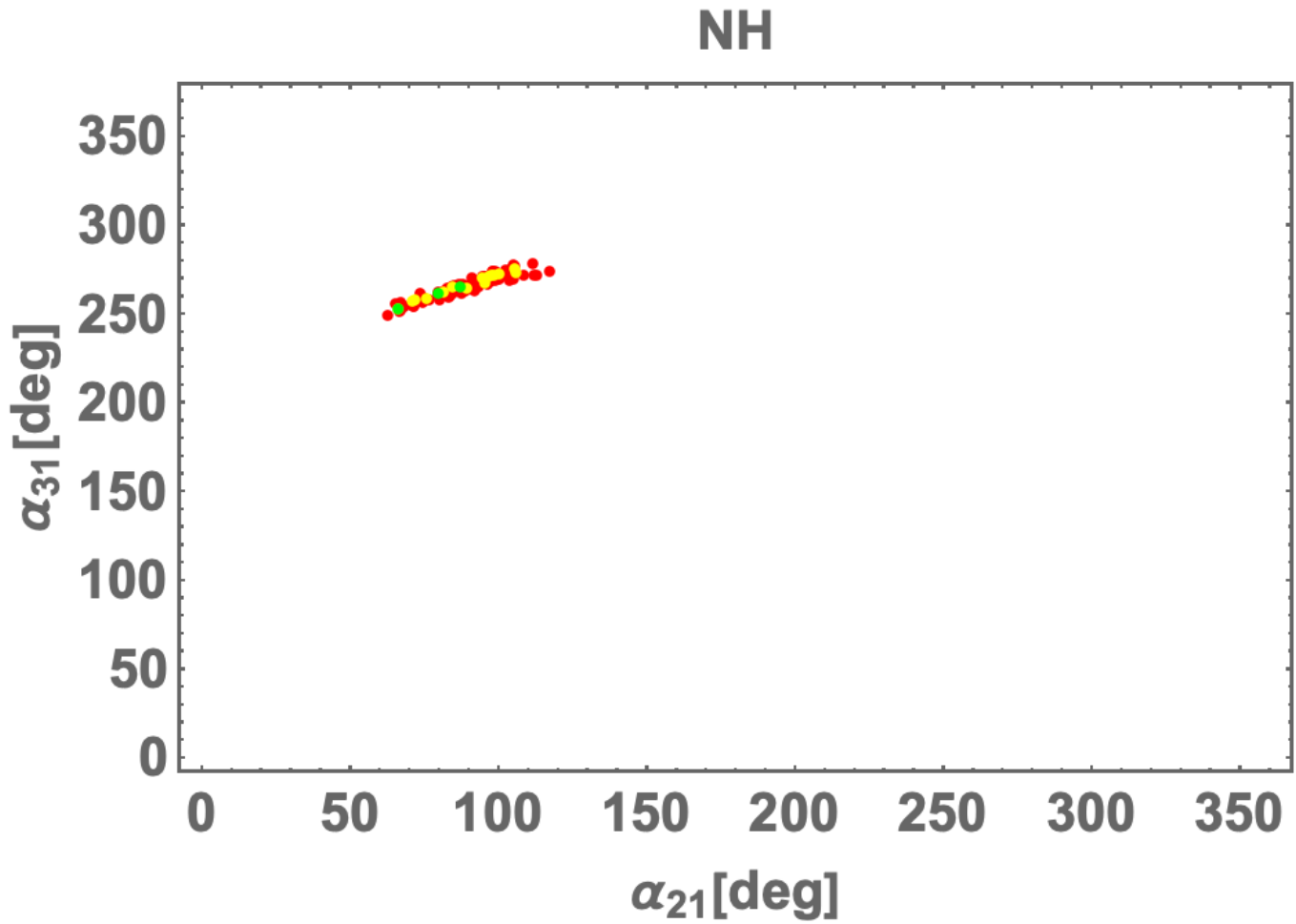}
  \caption{Allowed region for  Majorana phases meV in NH.}
  \label{fig:majo_nh}
\end{figure}
In Fig.~\ref{fig:majo_nh}, we show the allowed  region for Majorana phases and 
find $\alpha_{31} =[240-280]$ deg and $\alpha_{21} =[60-120]$ deg.
\begin{figure}[t]
    \includegraphics[width=77mm]{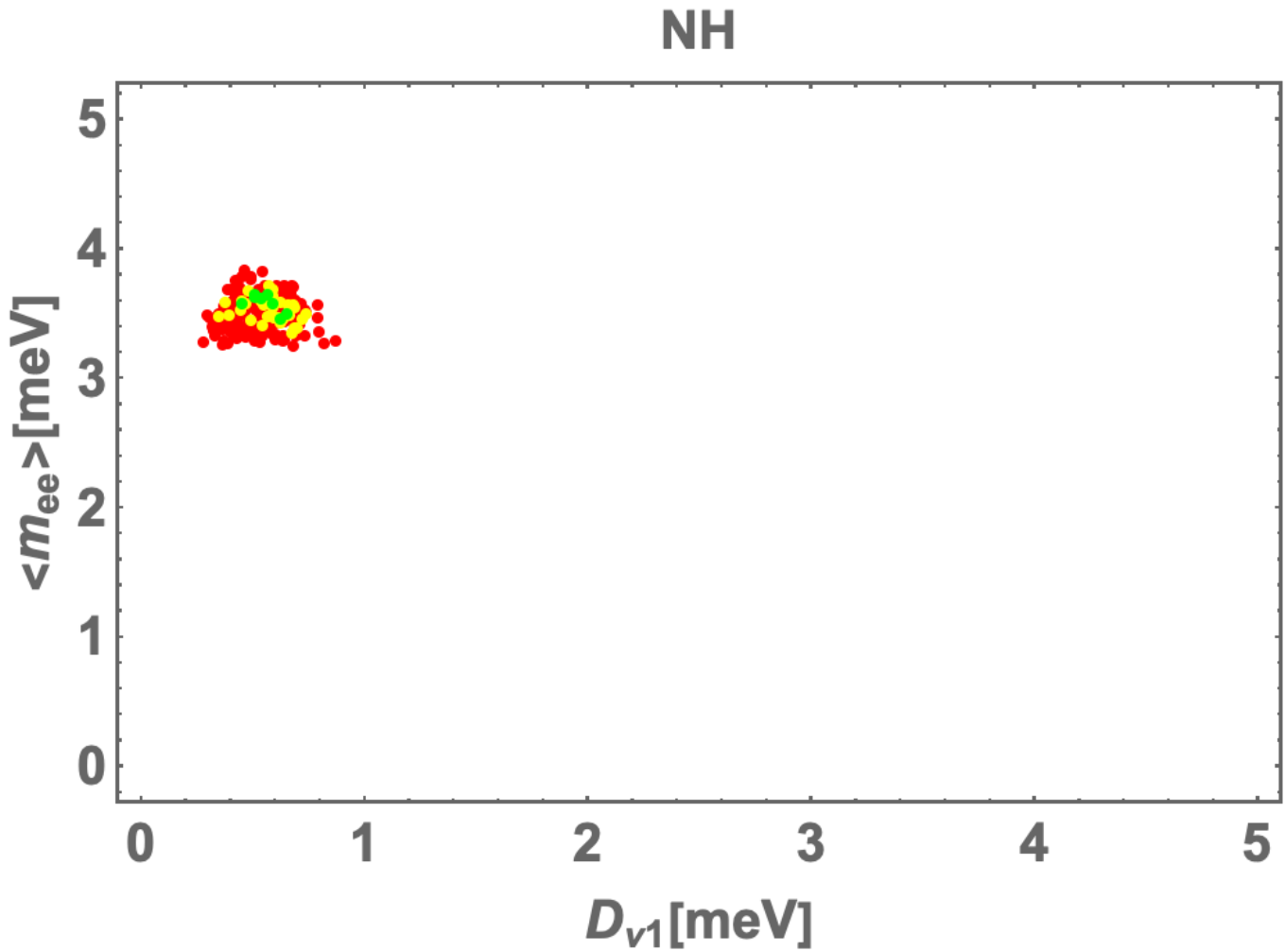}
  \caption{Allowed region for $\langle m_{ee}\rangle$ in terms of the lightest active neutrino mass in NH.}
  \label{fig:masses2_nh}
\end{figure}
In addition, in Fig.~\ref{fig:masses2_nh}, we show the allowed  region for $\langle m_{ee}\rangle$ in terms of the lightest active neutrino mass in NH and 
find the lightest active neutrino mass to be  $D_{\nu_1} =[0.2-1]$ meV.

\subsection{IH}


\begin{figure}[t]
  \includegraphics[width=77mm]{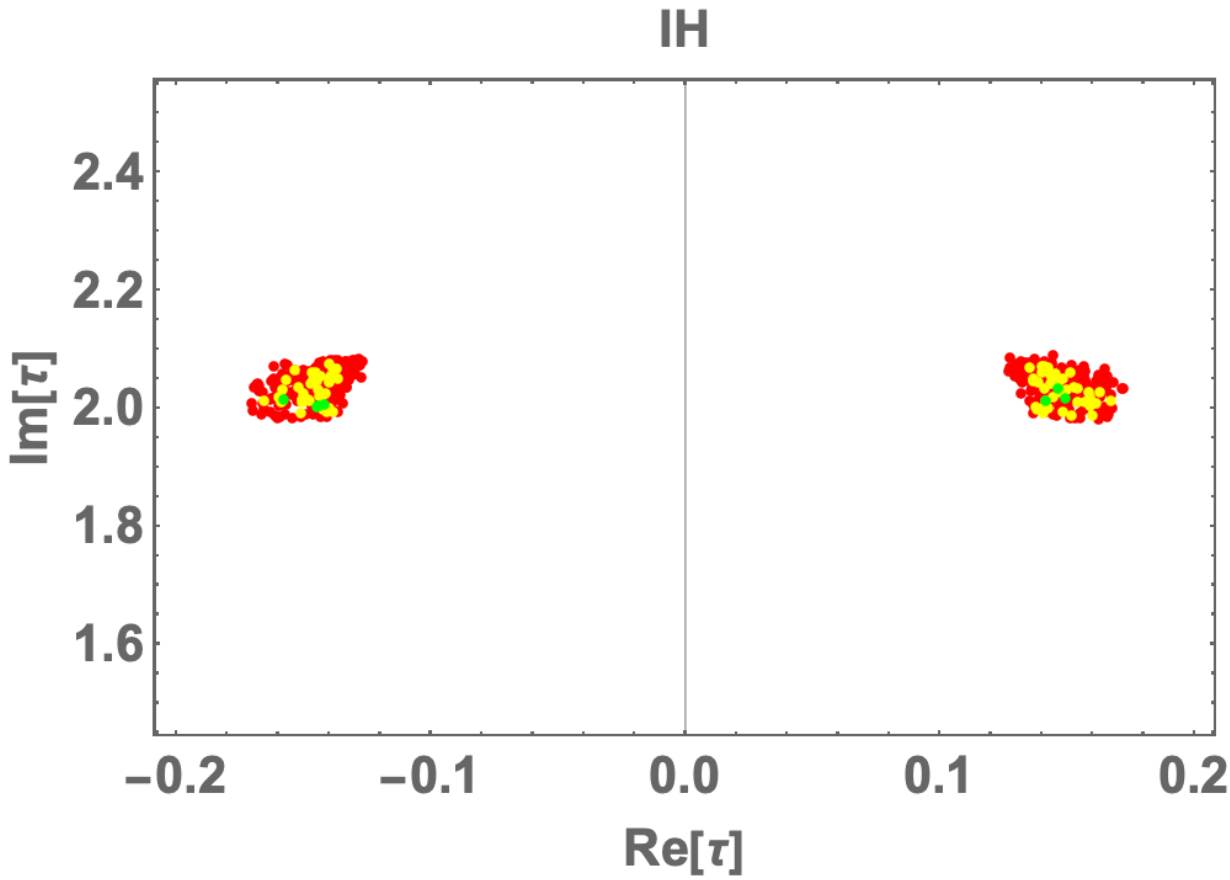}
  \caption{Allowed region for real $\tau$ and imaginary $\tau$ in IH.}
  \label{fig:tau_ih}
\end{figure}
In Fig.~\ref{fig:tau_ih}, we show the allowed region of $\tau$, and find that the allowed region is located at nearby $|{\rm Re}\tau|=[0.12-0.18]$ and ${\rm Im}\tau=[1.95-2.10]$.
\begin{figure}[t]
  \includegraphics[width=77mm]{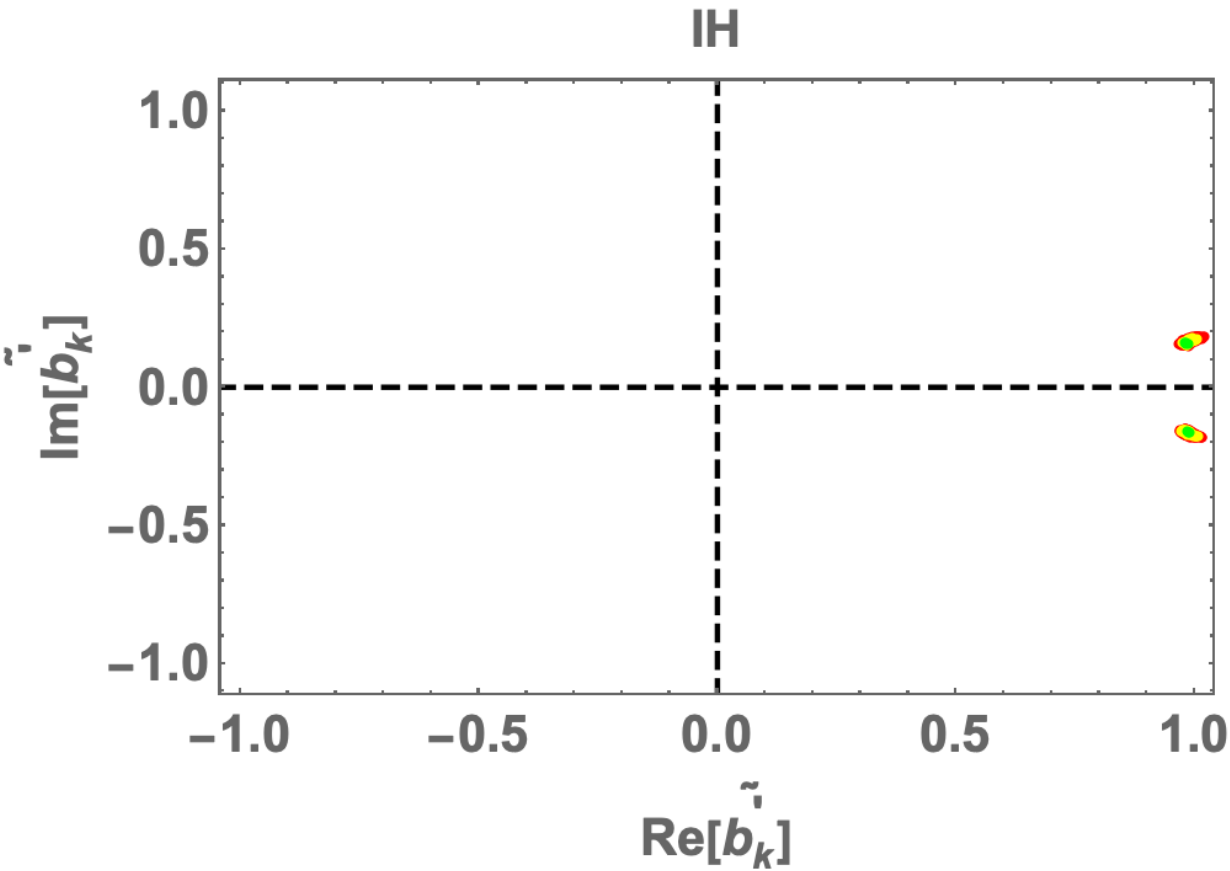}
  \caption{Allowed region for real $\tilde b'_k$ and imaginary $\tilde b'_k$ in IH.}
  \label{fig:bk_ih}
\end{figure}
In Fig.~\ref{fig:bk_ih}, we show the allowed region of $\tilde b'_k$, and find that the allowed region is about
${\rm Re}[\tilde b'_k]\approx1$ and $|{\rm Im}[\tilde b'_k]|=[0.1-0.2]$.

\begin{figure}[t]
  \includegraphics[width=77mm]{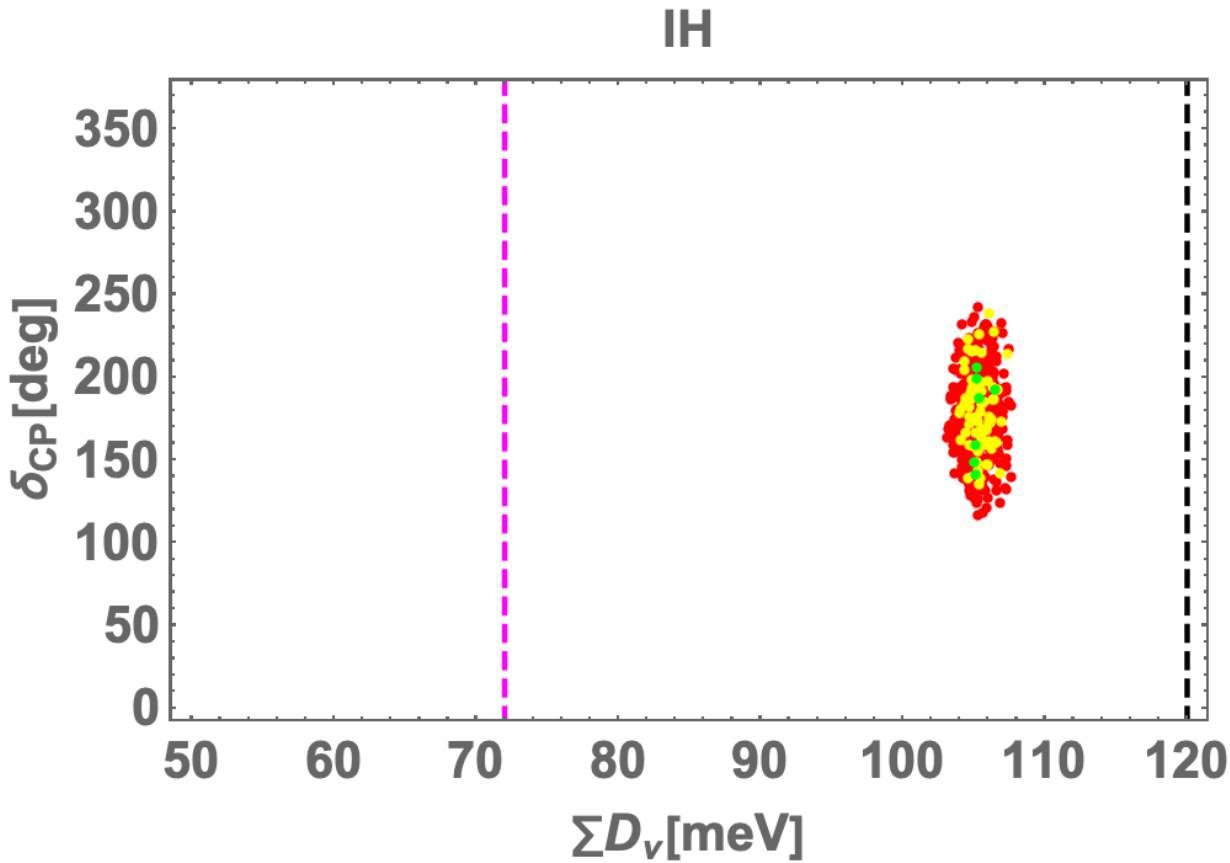}
    \includegraphics[width=77mm]{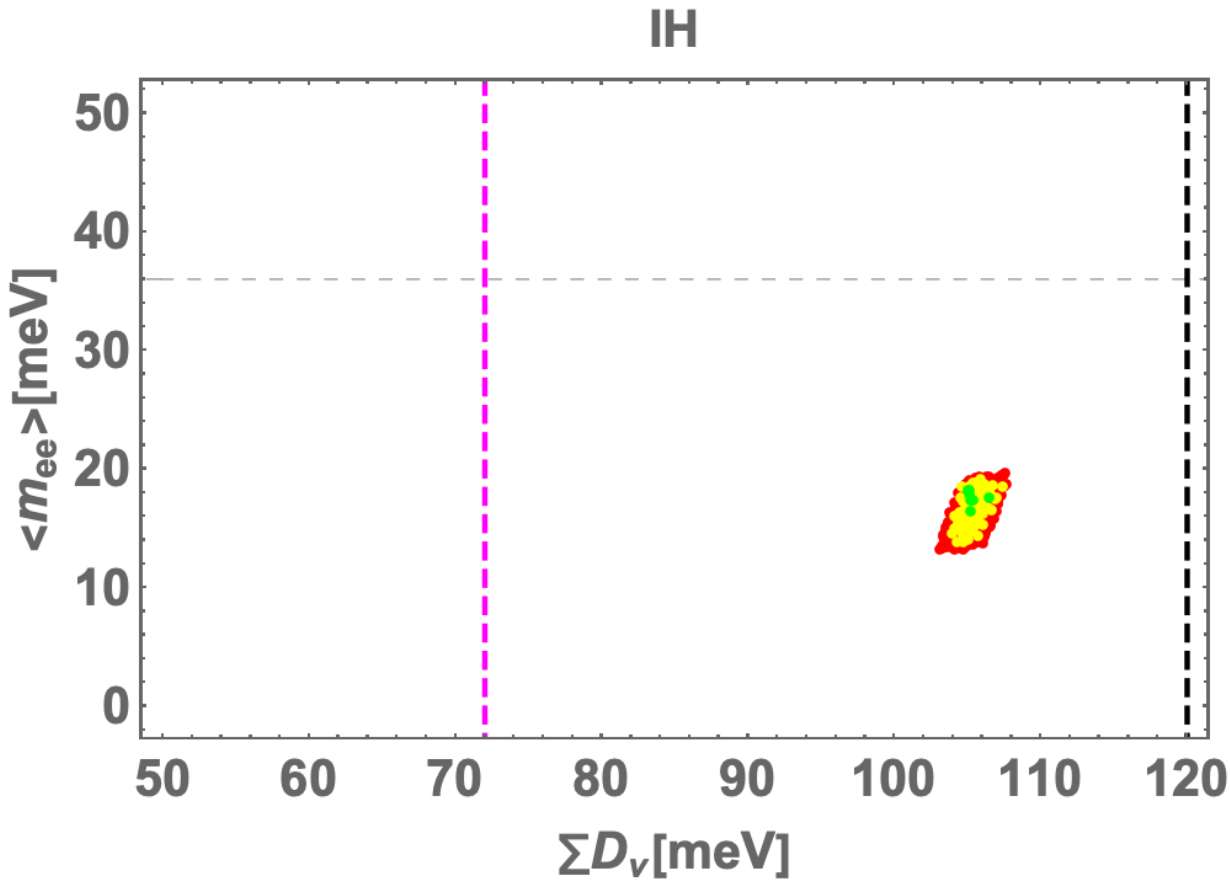}
  \caption{Allowed regions for  $\delta_{CP}$ deg (left) and $\langle m_{ee}\rangle$ meV (right) in terms of $\sum D_\nu$ meV in IH. The vertical magenta and black dotted lines are respectively upper bound on results of Planck+DESI~\cite{DESI:2024mwx}; $\sum D_\nu\le $72 meV and the minimal cosmological model $\Lambda$CDM $+\sum D_{\nu}$~\cite{Vagnozzi:2017ovm, Planck:2018vyg}; $\sum D_{\nu}\le$ 120 meV. The horizontal gray dotted line is lower bound on KamLAND-Zen data 36 meV. }
  \label{fig:sum-dcp_ih}
\end{figure}
In Fig.~\ref{fig:sum-dcp_ih}, we show the allowed  region for $\delta_{CP}$ deg (left) and $\langle m_{ee}\rangle$ meV (right) in terms of $\sum D_\nu$ meV and 
find $\delta_{CP} =[120-250]$ deg,  $\sum D_\nu =[102-108]$ meV, and $\langle m_{ee}\rangle =[12-20]$ meV.
The vertical magenta and black dotted lines are respectively upper bound on results of Planck+DESI~\cite{DESI:2024mwx}; $\sum D_\nu\le $72 meV and the minimal cosmological model $\Lambda$CDM $+\sum D_{\nu}$~\cite{Vagnozzi:2017ovm, Planck:2018vyg}; $\sum D_{\nu}\le$ 120 meV.
The horizontal gray dotted line is the lower bound on the KamLAND-Zen data 36 meV. 

\begin{figure}[t]
  \includegraphics[width=77mm]{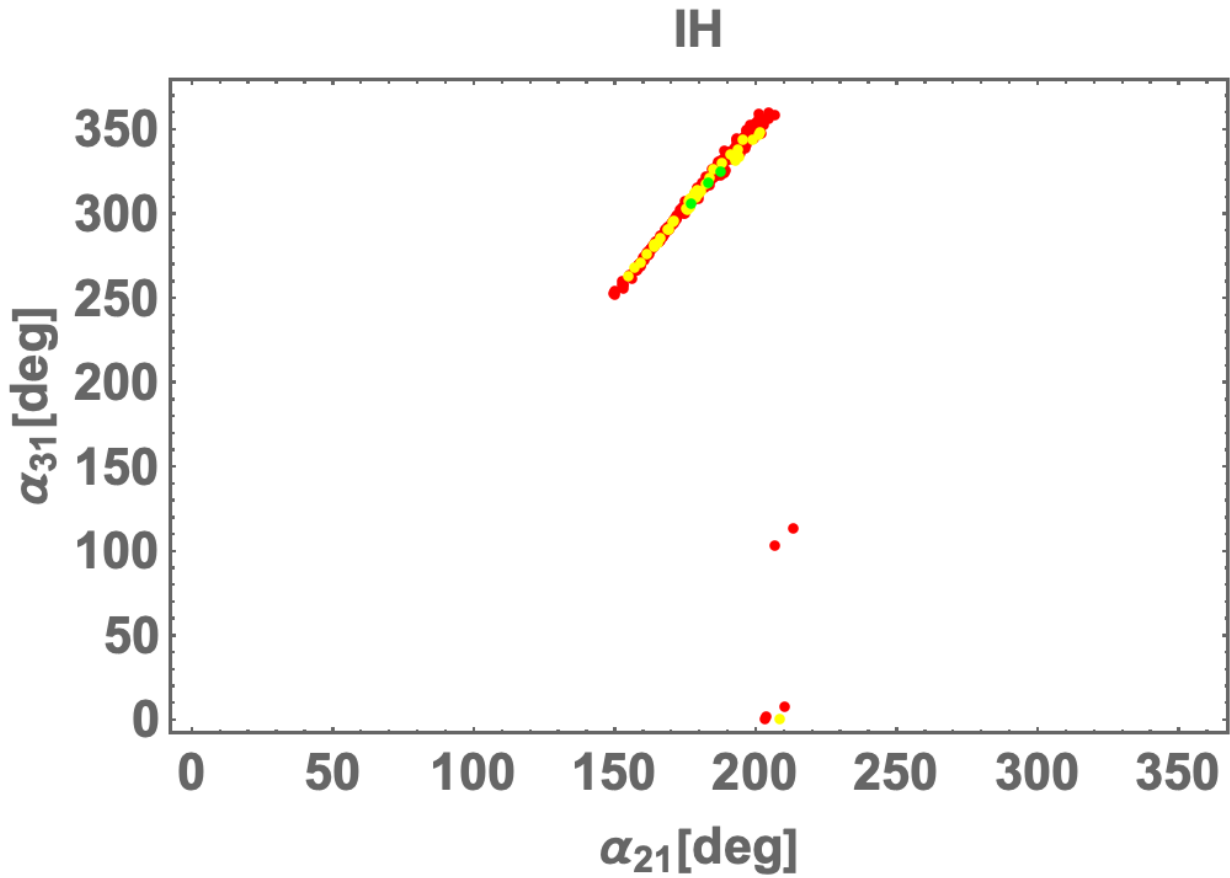}
  \caption{Allowed region for  Majorana phases meV in IH.}
  \label{fig:majo_ih}
\end{figure}
In Fig.~\ref{fig:majo_ih}, we show the allowed region for Majorana phases and
find $\alpha_{31} =[0-10,\ 100-120,\ 240-360]$ deg and $\alpha_{21} =[140-210]$ deg.
\begin{figure}[t]
    \includegraphics[width=77mm]{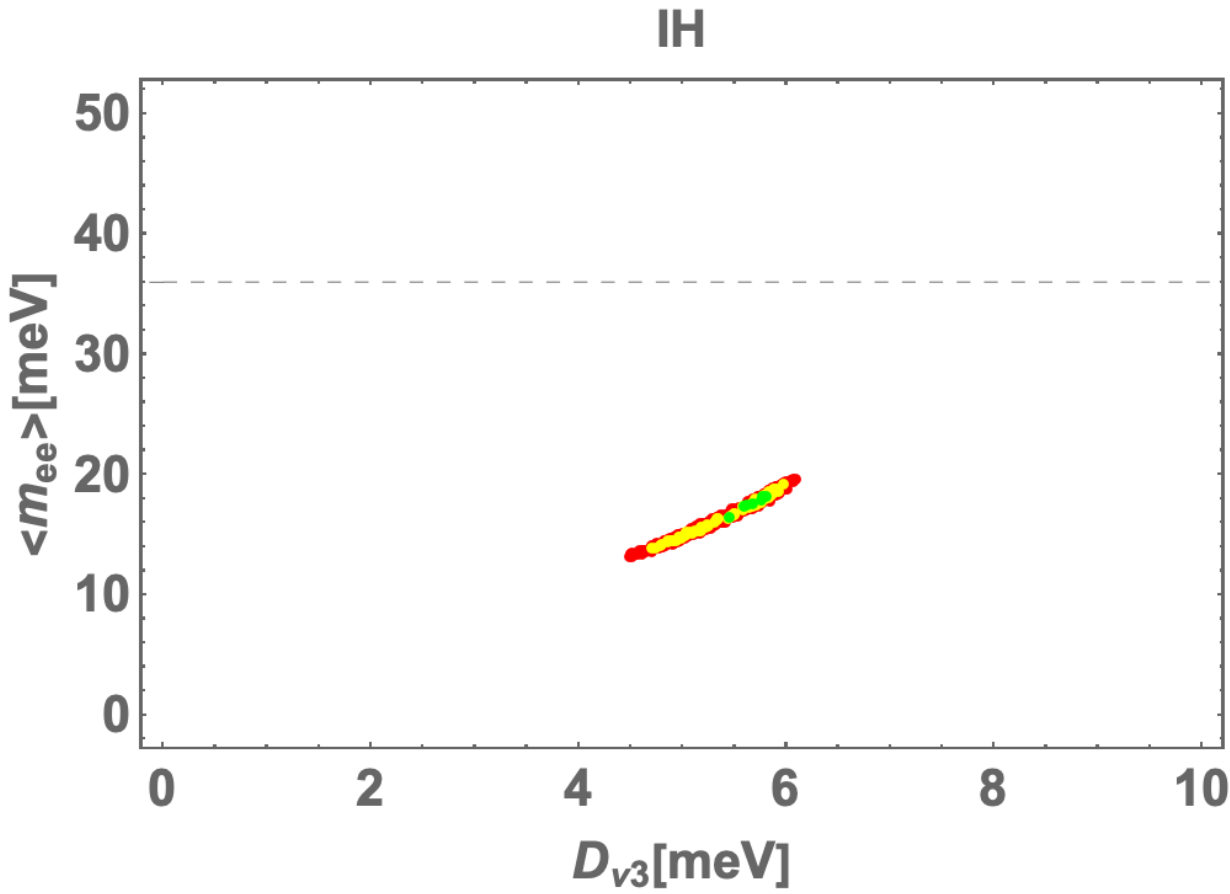}
  \caption{Allowed region for $\langle m_{ee}\rangle$ in terms of the lightest active neutrino mass in IH.}
  \label{fig:masses2_ih}
\end{figure}
In Fig.~\ref{fig:masses2_ih}, we show the allowed  region for $\langle m_{ee}\rangle$ in terms of the lightest active neutrino mass in IH and 
find the lightest active neutrino mass to be  $D_{\nu_3} =[4.5-6.2]$ meV.

{
\section{Analytical evaluations}
{In numerical analysis, we obtain clustering of resulting points since the number of free parameters is few and the range of allowed parameters is constrained by observed data. In fact we only have two complex parameters $\tau$ and $\tilde{b}'_k$ for realizing neutrino mixing angles and $\Delta m^2_{\rm sol}$ ($\Delta m^2_{\rm atm}$ is fitted by $\kappa$).
To understand the behavior, we discuss neutrino mass matrix analytically. }
Since our allowed region in case of NH is localized at nearby a fixed point of $\omega\equiv e^{2\pi i/3}$, 
analytical estimation would be possible.~\footnote{In general, it is difficult to evaluate analytical evaluations of $\tau$ except fixed points. Thus, we would not deal with the case of IH even though the allowed region is localized like the case of NH.}
At nearby $\tau=\omega$, we expand $Y^{(0)}_3$~\cite{Kobayashi:2025hnc} and $Y^{(2)}_3$~\cite{Okada:2020ukr} at this point as follows:
\begin{align}
y_1&\simeq -\kappa_0\frac{{\rm Re}[\omega]}{{\rm Im}[\omega]} + \kappa_1 \omega^*(\epsilon-\epsilon^*) + {\cal O}(\epsilon^2),\\
y_2&\simeq -\kappa_0\frac{\omega}{{\rm Im}[\omega]} + \kappa_1 \omega^*\left(\epsilon+\frac12 \epsilon^*\right)+ {\cal O}(\epsilon^2),\\
y_3&\simeq -\kappa_0\frac{\omega^*}{{\rm Im}[\omega]} - \kappa_1 \omega^*\left(\frac12 \epsilon+ \epsilon^*\right)+ {\cal O}(\epsilon^2),\\
f_1&\simeq Y_0,\\
f_2&\simeq Y_0 \left(\omega+\epsilon_1 \right)+ {\cal O}(\epsilon_1^2),\\
f_3&\simeq -Y_0 \frac{\omega^2}{2} \left(1+\epsilon_2 \right)+ {\cal O}(\epsilon_2^2),
\end{align} 
where $\kappa_0\approx $0.151, $\kappa_1\approx$0.411 + 0.237 $i$, $Y_0\approx 0.949$, $\epsilon_1=\epsilon_2/2 =2.1 i\epsilon$, and 
$\epsilon$ is a small deviation from $\tau=\omega$.
Applying the above expansions, the mass matrix of charged-lepton at leading order can be diagonalized by the following mixing matrices:
\begin{align}
&V_{eL}^0 = \frac13
\begin{pmatrix}
2 & 2 & -1 \\ 
 -\omega^* & 2\omega^* &2\omega^* \\ 
2\omega & -\omega & 2\omega \\ 
\end{pmatrix} +{\cal O}(\epsilon),\quad
 V_{eR}^0 = \frac13
\begin{pmatrix}
-2 & 2 & -1 \\ 
-2\omega & -\omega & 2\omega \\ 
 \omega^* & 2\omega^* &2\omega^* \\ 
\end{pmatrix}+{\cal O}(\epsilon) ,
 \label{eq:cglp-lead}
\end{align}
 and we find the mass eigenvalues as 
 \begin{align}
&m_{e}^0 = 
-a_e +\sqrt3 \kappa_0(-b_e+c_e)+{\cal O}(\epsilon) ,\quad
 m_{\mu}^0 = 
a_e +\sqrt3 \kappa_0(b_e+c_e) +{\cal O}(\epsilon),\quad
 m_{\tau}^0 = 
a_e -2\sqrt3 \kappa_0b_e+{\cal O}(\epsilon) .
 \label{eq:cglp-lead}
\end{align}
 Now that the charged-lepton sector can be written in terms of analytical form as the leading order, we can also find the mass matrix of neutrinos:
 \begin{align}
m_\nu^0& \sim 
\begin{pmatrix}
 -2 (4 m_e  - b_m m_\mu) & -\frac12 (16 m_e+5 b_m m_\mu)\omega^* & \frac13 (16 m_e - 5 b_m m_\mu -4 b_p m_\tau)\omega \\ 
 -\frac12 (16 m_e + 5 b_m m_\mu)\omega^* &  -2 (4 m_e  - b_m m_\mu)\omega & \frac13 (16 m_e + b_m m_\mu +2 b_p m_\tau) \\ 
 \frac13 (16 m_e - 5 b_m m_\mu -4 b_p m_\tau)\omega & \frac13 (16 m_e + b_m m_\mu +2 b_p m_\tau)  & -\frac23 (5 m_e + 2 b_m m_\mu+ 4 b_p m_\tau)\omega^*  \\ 
\end{pmatrix} +{\cal O}(\epsilon) \label{eq:approxneut1}
,
\end{align}
 where $b_m\equiv -1+\tilde b'_k$ and $b_p\equiv 1+\tilde b'_k$.
On the other hand, the neutrino mass matrix at the leading order $m^0_\nu$ can be written in terms of observables and some parameters via definitions $m_\nu=U^*_\nu D_\nu U\dag_\nu$ and $U_{PMNS}\equiv V_{eL}^\dag U_\nu$:
 \begin{align}
 m^0_\nu \sim
 V^{0*}_{eL} U^*_{PMNS} D_\nu U^\dag_{PMNS} V^{0\dag}_{eL} 
    +{\cal O}(\epsilon).
\end{align}
The above relation is explicitly given in both NH and IH cases such that 
 \begin{align}
({\rm NH}): & ~\nn\\
(m^0_\nu)_{11}& \sim e^{i\alpha_{31}}m_3(c_{23}-2s_{23})^2+e^{i\alpha_{21}}m_2(2s_{12}+c_{12}(2c_{23}+s_{23}))^2 +{\cal O}(\epsilon), \label{eq:approxneutNH1} \\
(m^0_\nu)_{12}&\sim \left[-2 e^{i\alpha_{31}}m_3(c_{23}-2s_{23})(c_{23}+s_{23}) 
-e^{\alpha_{21}}m_2(s_{12}+2c_{12}(-c_{23}+s_{23}))(2s_{12}+c_{12}(2c_{23}+s_{23})) \right]\omega\nn\\
&+{\cal O}(\epsilon), \\
(m^0_\nu)_{13}&\sim \left[- e^{i\alpha_{31}}m_3(c_{23}-2s_{23})(2c_{23}-s_{23}) 
+e^{\alpha_{21}}m_2(2s_{12}+c_{12}(2c_{23}+s_{23}))(2s_{12}-c_{12}(c_{23}+2s_{23})) \right]\omega^* \nn\\
&+{\cal O}(\epsilon), \\
(m^0_\nu)_{22}&\sim 4 e^{i\alpha_{31}}m_3(c_{23}+s_{23})^2+e^{i\alpha_{21}}m_2(s_{12}+2c_{12}(-c_{23}+s_{23}))^2\omega^2 +{\cal O}(\epsilon), \\
(m^0_\nu)_{23}&\sim \left[2 e^{i\alpha_{31}}m_3(2c_{23}-s_{23})(c_{23}+s_{23}) 
-e^{\alpha_{21}}m_2(s_{12}+2c_{12}(-c_{23}+s_{23}))(2s_{12}-c_{12}(c_{23}+2s_{23})) \right] 
 \nn\\
&+{\cal O}(\epsilon), \\
(m^0_\nu)_{33}&\sim  e^{i\alpha_{31}}m_3(-2c_{23}+s_{23})^2+e^{i\alpha_{21}}m_2(-2s_{12}+c_{12}(c_{23}+2s_{23}))^2\omega^{*2} +{\cal O}(\epsilon), 
\label{eq:approxneutNH}
\end{align}
and
 \begin{align}
 ({\rm IH}): & \nn\\
(m^0_\nu)_{11}&\sim e^{i\alpha_{21}}m_2(2s_{12}+c_{12}(2c_{23}+s_{23}))^2+m_1(-2c_{12}+s_{12}(2c_{23}+s_{23}))^2 +{\cal O}(\epsilon), \label{eq:approxneutIH1} \\
(m^0_\nu)_{12}&\sim \left[- e^{i\alpha_{21}}m_2(s_{12}+2c_{12}(-c_{23}+s_{23}))(2s_{12}+c_{12}(2c_{23}+s_{23})) \right. \nn\\
&\left. \hspace{1.7cm}
- m_1(c_{12}+2s_{12}(c_{23}-s_{23}))(2c_{12}-s_{12}(2c_{23}+s_{23})) \right]\omega  +{\cal O}(\epsilon), \\
(m^0_\nu)_{13}&\sim \left[ e^{i\alpha_{21}}m_2(2s_{12}+c_{12}(2c_{23}+s_{23}))(2s_{12}-c_{12}(c_{23}+2s_{23})) \right. \nn\\
&\left.\hspace{1.7cm}
+m_1(2c_{12}-s_{12}(2c_{23}+s_{23}))(2c_{12}+s_{12}(c_{23}+2s_{23})) \right]\omega^* +{\cal O}(\epsilon), \\
(m^0_\nu)_{22}&\sim m_1(c_{12}+2s_{12}(c_{23}-s_{23})^2+e^{i\alpha_{21}}m_2(s_{12}+2c_{12}(-c_{23}+s_{23}))^2\omega^2 +{\cal O}(\epsilon), \\
(m^0_\nu)_{23}&\sim \left[- e^{i\alpha_{21}}m_2(s_{12}+2c_{12}(-c_{23}+s_{23}))(2s_{12}-c_{12}(c_{23}+2s_{23})) \right. \nn\\
&\left. \hspace{1.7cm}
-m_1(c_{12}+2s_{12}(c_{23}-s_{23}))(2c_{12}+s_{12}(c_{23}+2s_{23})) \right] 
+{\cal O}(\epsilon), \\
(m^0_\nu)_{33}&\sim  e^{i\alpha_{21}}m_2(-2s_{12}+c_{12}(c_{23}+2s_{23}))^2+m_1(2c_{12}+s_{12}(c_{23}+2s_{23}))^2\omega^{*2} +{\cal O}(\epsilon),\label{eq:approxneutIH}
\end{align}
 where we have ignored the lightest neutrino mass. 
{Comparing Eq.~(\ref{eq:approxneut1}) with  Eq.~(\ref{eq:approxneutNH1})-(\ref{eq:approxneutIH}) for NH and Eq.~(\ref{eq:approxneutIH1})-(\ref{eq:approxneutIH}) for IH,
 we can check if our theoretical input parameter $\tilde b'_k$ can be consistent with observed data.
 As a result, we can find the solutions of $\tilde b'_k$ consistent with the data in NH case while there is no solution for IH case for $\tau \sim \omega$. 
 In this way the allowed range of our free parameters is constrained since neutrino mass matrix is given by $\tau$ and $\tilde b'_k$ and only limited range can be allowed.
 }
}


\section{Conclusions and discussions}
We have investigated a Zee model applying a non-holomorphic modular $A_4$ symmetry, 
and we have obtained unique and sharp predictions for each of the normal and inverted hierarchy.
This is because we have only two parameters in the neutrino sector including modulus $\tau$
due to appropriately assuming that the masses of charged-leptons are negligibly less than the masses for singly charged
bosons.

Before closing this paper, we mention the constraints of Yukawa couplings $f$ and $y$.
These main constraints come from lepton/hadron universalities which are induced at tree level and charged lepton flavor violations
which are induced at one-loop level~\cite{Herrero-Garcia:2014hfa, Okada:2015nga, Lindner:2016bgg}.
Even when we consider these bounds, it is totally safe if we take these Yukawa couplings are 0.01 in case of $m_{s}\sim m_{h'}=$1 TeV.
This value is easily achieved by adjusting the overall factor $a'$ and $a_s$, which are parts of $\kappa$.
Although $\kappa$ is fixed to fit the atmospheric mass squared difference, $\kappa$ can still maintain the fit value due to
changing $\mu$ and $r$.

\section*{Acknowledgments}
\vspace{0.5cm}
The work was supported by the Fundamental Research Funds for the Central Universities (T.~N.). 


\bibliography{NonHoloMa4_zee.bib}

\begin{thebibliography}{16}
\expandafter\ifx\csname natexlab\endcsname\relax\def\natexlab#1{#1}\fi
\expandafter\ifx\csname bibnamefont\endcsname\relax
  \def\bibnamefont#1{#1}\fi
\expandafter\ifx\csname bibfnamefont\endcsname\relax
  \def\bibfnamefont#1{#1}\fi
\expandafter\ifx\csname citenamefont\endcsname\relax
  \def\citenamefont#1{#1}\fi
\expandafter\ifx\csname url\endcsname\relax
  \def\url#1{\texttt{#1}}\fi
\expandafter\ifx\csname urlprefix\endcsname\relax\def\urlprefix{URL }\fi
\providecommand{\bibinfo}[2]{#2}
\providecommand{\eprint}[2][]{\url{#2}}

\bibitem[{\citenamefont{Qu and Ding}(2024)}]{Qu:2024rns}
\bibinfo{author}{\bibfnamefont{B.-Y.} \bibnamefont{Qu}} \bibnamefont{and}
  \bibinfo{author}{\bibfnamefont{G.-J.} \bibnamefont{Ding}},
  \bibinfo{journal}{JHEP} \textbf{\bibinfo{volume}{08}}, \bibinfo{pages}{136}
  (\bibinfo{year}{2024}), \eprint{2406.02527}.

\bibitem[{\citenamefont{Ding et~al.}(2024)\citenamefont{Ding, Lu, Petcov, and
  Qu}}]{Ding:2024inn}
\bibinfo{author}{\bibfnamefont{G.-J.} \bibnamefont{Ding}},
  \bibinfo{author}{\bibfnamefont{J.-N.} \bibnamefont{Lu}},
  \bibinfo{author}{\bibfnamefont{S.~T.} \bibnamefont{Petcov}},
  \bibnamefont{and} \bibinfo{author}{\bibfnamefont{B.-Y.} \bibnamefont{Qu}}
  (\bibinfo{year}{2024}), \eprint{2408.15988}.

\bibitem[{\citenamefont{Li et~al.}(2024)\citenamefont{Li, Lu, and
  Ding}}]{Li:2024svh}
\bibinfo{author}{\bibfnamefont{C.-C.} \bibnamefont{Li}},
  \bibinfo{author}{\bibfnamefont{J.-N.} \bibnamefont{Lu}}, \bibnamefont{and}
  \bibinfo{author}{\bibfnamefont{G.-J.} \bibnamefont{Ding}}
  (\bibinfo{year}{2024}), \eprint{2410.24103}.

\bibitem[{\citenamefont{Nomura and Okada}(2024)}]{Nomura:2024atp}
\bibinfo{author}{\bibfnamefont{T.}~\bibnamefont{Nomura}} \bibnamefont{and}
  \bibinfo{author}{\bibfnamefont{H.}~\bibnamefont{Okada}}
  (\bibinfo{year}{2024}), \eprint{2408.01143}.

\bibitem[{\citenamefont{Nomura et~al.}(2025)\citenamefont{Nomura, Okada, and
  Popov}}]{Nomura:2024vzw}
\bibinfo{author}{\bibfnamefont{T.}~\bibnamefont{Nomura}},
  \bibinfo{author}{\bibfnamefont{H.}~\bibnamefont{Okada}}, \bibnamefont{and}
  \bibinfo{author}{\bibfnamefont{O.}~\bibnamefont{Popov}},
  \bibinfo{journal}{Phys. Lett. B} \textbf{\bibinfo{volume}{860}},
  \bibinfo{pages}{139171} (\bibinfo{year}{2025}), \eprint{2409.12547}.

\bibitem[{\citenamefont{Zee}(1980)}]{Zee:1980ai}
\bibinfo{author}{\bibfnamefont{A.}~\bibnamefont{Zee}}, \bibinfo{journal}{Phys.
  Lett. B} \textbf{\bibinfo{volume}{93}}, \bibinfo{pages}{389}
  (\bibinfo{year}{1980}), \bibinfo{note}{[Erratum: Phys.Lett.B 95, 461
  (1980)]}.

\bibitem[{\citenamefont{Ma}(2006)}]{Ma:2006km}
\bibinfo{author}{\bibfnamefont{E.}~\bibnamefont{Ma}}, \bibinfo{journal}{Phys.
  Rev. D} \textbf{\bibinfo{volume}{73}}, \bibinfo{pages}{077301}
  (\bibinfo{year}{2006}), \eprint{hep-ph/0601225}.

\bibitem[{\citenamefont{Nomura et~al.}(2021)\citenamefont{Nomura, Okada, and
  Qi}}]{Nomura:2021pld}
\bibinfo{author}{\bibfnamefont{T.}~\bibnamefont{Nomura}},
  \bibinfo{author}{\bibfnamefont{H.}~\bibnamefont{Okada}}, \bibnamefont{and}
  \bibinfo{author}{\bibfnamefont{Y.-h.} \bibnamefont{Qi}}
  (\bibinfo{year}{2021}), \eprint{2111.10944}.

\bibitem[{\citenamefont{Abe et~al.}(2024)}]{KamLAND-Zen:2024eml}
\bibinfo{author}{\bibfnamefont{S.}~\bibnamefont{Abe}} \bibnamefont{et~al.}
  (\bibinfo{collaboration}{KamLAND-Zen}) (\bibinfo{year}{2024}),
  \eprint{2406.11438}.

\bibitem[{\citenamefont{Vagnozzi et~al.}(2017)\citenamefont{Vagnozzi, Giusarma,
  Mena, Freese, Gerbino, Ho, and Lattanzi}}]{Vagnozzi:2017ovm}
\bibinfo{author}{\bibfnamefont{S.}~\bibnamefont{Vagnozzi}},
  \bibinfo{author}{\bibfnamefont{E.}~\bibnamefont{Giusarma}},
  \bibinfo{author}{\bibfnamefont{O.}~\bibnamefont{Mena}},
  \bibinfo{author}{\bibfnamefont{K.}~\bibnamefont{Freese}},
  \bibinfo{author}{\bibfnamefont{M.}~\bibnamefont{Gerbino}},
  \bibinfo{author}{\bibfnamefont{S.}~\bibnamefont{Ho}}, \bibnamefont{and}
  \bibinfo{author}{\bibfnamefont{M.}~\bibnamefont{Lattanzi}},
  \bibinfo{journal}{Phys. Rev. D} \textbf{\bibinfo{volume}{96}},
  \bibinfo{pages}{123503} (\bibinfo{year}{2017}), \eprint{1701.08172}.

\bibitem[{\citenamefont{Aghanim et~al.}(2020)}]{Planck:2018vyg}
\bibinfo{author}{\bibfnamefont{N.}~\bibnamefont{Aghanim}} \bibnamefont{et~al.}
  (\bibinfo{collaboration}{Planck}), \bibinfo{journal}{Astron. Astrophys.}
  \textbf{\bibinfo{volume}{641}}, \bibinfo{pages}{A6} (\bibinfo{year}{2020}),
  \bibinfo{note}{[Erratum: Astron.Astrophys. 652, C4 (2021)]},
  \eprint{1807.06209}.

\bibitem[{\citenamefont{Adame et~al.}(2024)}]{DESI:2024mwx}
\bibinfo{author}{\bibfnamefont{A.~G.} \bibnamefont{Adame}} \bibnamefont{et~al.}
  (\bibinfo{collaboration}{DESI}) (\bibinfo{year}{2024}), \eprint{2404.03002}.

\bibitem[{\citenamefont{Esteban et~al.}(2024)\citenamefont{Esteban,
  Gonzalez-Garcia, Maltoni, Martinez-Soler, Pinheiro, and
  Schwetz}}]{Esteban:2024eli}
\bibinfo{author}{\bibfnamefont{I.}~\bibnamefont{Esteban}},
  \bibinfo{author}{\bibfnamefont{M.~C.} \bibnamefont{Gonzalez-Garcia}},
  \bibinfo{author}{\bibfnamefont{M.}~\bibnamefont{Maltoni}},
  \bibinfo{author}{\bibfnamefont{I.}~\bibnamefont{Martinez-Soler}},
  \bibinfo{author}{\bibfnamefont{J.~a.~P.} \bibnamefont{Pinheiro}},
  \bibnamefont{and} \bibinfo{author}{\bibfnamefont{T.}~\bibnamefont{Schwetz}}
  (\bibinfo{year}{2024}), \eprint{2410.05380}.

\bibitem[{\citenamefont{Herrero-Garcia
  et~al.}(2014)\citenamefont{Herrero-Garcia, Nebot, Rius, and
  Santamaria}}]{Herrero-Garcia:2014hfa}
\bibinfo{author}{\bibfnamefont{J.}~\bibnamefont{Herrero-Garcia}},
  \bibinfo{author}{\bibfnamefont{M.}~\bibnamefont{Nebot}},
  \bibinfo{author}{\bibfnamefont{N.}~\bibnamefont{Rius}}, \bibnamefont{and}
  \bibinfo{author}{\bibfnamefont{A.}~\bibnamefont{Santamaria}},
  \bibinfo{journal}{Nucl. Phys. B} \textbf{\bibinfo{volume}{885}},
  \bibinfo{pages}{542} (\bibinfo{year}{2014}), \eprint{1402.4491}.

\bibitem[{\citenamefont{Okada}(2015)}]{Okada:2015nga}
\bibinfo{author}{\bibfnamefont{H.}~\bibnamefont{Okada}} (\bibinfo{year}{2015}),
  \eprint{1503.04557}.

\bibitem[{\citenamefont{Lindner et~al.}(2018)\citenamefont{Lindner, Platscher,
  and Queiroz}}]{Lindner:2016bgg}
\bibinfo{author}{\bibfnamefont{M.}~\bibnamefont{Lindner}},
  \bibinfo{author}{\bibfnamefont{M.}~\bibnamefont{Platscher}},
  \bibnamefont{and} \bibinfo{author}{\bibfnamefont{F.~S.}
  \bibnamefont{Queiroz}}, \bibinfo{journal}{Phys. Rept.}
  \textbf{\bibinfo{volume}{731}}, \bibinfo{pages}{1} (\bibinfo{year}{2018}),
  \eprint{1610.06587}.

\end{thebibliography}

\end{document}